\documentclass[11pt]{article}

\usepackage[english]{babel}    

\usepackage{epsfig,amsfonts,graphics,fullpage}
\usepackage{graphics,times,latexsym,amsfonts,amsmath}
\usepackage[sectionbib]{natbib}
\usepackage{epsfig}
\usepackage{xspace}
\usepackage{color}
\newtheorem{example}{Example}[section]
\begin{document}



\title {Towards a Statistical  Methodology to Evaluate Program Speedups and their Optimisation Techniques}
\author{Sid-Ahmed-Ali \textsc{Touati}\\\texttt{Sid.Touati@uvsq.fr}\\University of Versailles Saint-Quentin en Yvelines, France}

\date{April 2009}

\maketitle
\begin{abstract}
The community of program optimisation and analysis, code performance evaluation, parallelisation and optimising compilation has published since many decades hundreds of research and engineering articles in major conferences and journals. These articles study efficient algorithms, strategies and techniques to accelerate programs execution times, or optimise other performance metrics (MIPS, code size, energy/power, MFLOPS, etc.). Many speedups are published, but nobody is able to reproduce them exactly. The non-reproducibility of our research results is a dark point of the art, and we cannot be qualified as {\it computer scientists} if we do not provide rigorous experimental methodology. 

This article provides a first effort towards a correct statistical protocol for analysing and measuring speedups. As we will see, some common mistakes are done by the community inside published articles, explaining part of the non-reproducibility of the results. Our current article is not sufficient by its own to deliver a complete experimental methodology, further efforts must be done by the community to decide about a common protocol for our future experiences. Anyway, our community should take care about the aspect of reproducibility of the results in the future.
\end{abstract}

\section{Introduction}
The community of program optimisation and analysis, code performance evaluation, parallelisation and optimising compilation has published since many decades hundreds of research and engineering articles in major conferences and journals. These articles study efficient algorithms, strategies and techniques to accelerate programs execution times, or optimise other performance metrics (MIPS, code size, energy/power, MFLOPS, etc.). The efficiency of a code optimisation technique is generally published according to two principles, non necessarily disjoint. The first principle is to provide a mathematical proof given a theoretical model that the published research result is correct or/and efficient: this is the hard part of research in computer science, since if the model is too simple, it would not represent real world, and if the model is too close to real world, mathematics become too complex to digest. A second principle is to propose and implement a code optimisation technique and to practice it on a set of chosen benchmarks in order to evaluate its efficiency. This article concerns this last point: how can we convince the community by rigorous statistics that the experimental study publishes correct and fair results ?

\subsection{Non-Reproducible Experimental Results}
Hard natural sciences such as physics, chemistry and biology impose strict experimental methodologies and rigorous statistical measures in order to guarantee the reproducibility  of the results. The reproducibility of the experimental results in our community is, namely, our dark point. Given a research article, it is in practice impossible or too difficult to reproduce the published performance. If our results are not reproducible, we cannot say that we are doing science! Some aspects make a research article non-reproducible:
\begin{itemize}
	\item Non using precise scientific languages such as mathematics. Ideally, mathematics must always be preferred to describe ideas, if possible, with an accessible difficulty. 
	\item Non available software, non released software, non communicated precise data.
	\item Not providing formal algorithms or protocols make impossible to reproduce exactly the ideas. For instance, the authors in  \cite{temam:micro04} spent large efforts to re-implement some branch predictor algorithms based on the published research articles, but they fail to reproduce the initial results of the authors. Simply because the initial articles describing the branch predictors are not formal, so they can be interpreted differently.
	\item Hide many experimental details. As demonstrated by \cite{MDSH09}, bringing small modification on the execution environment brings contradictory experimental results. For instance, just changing the size of the linux shell variables or the order of linking an application alter the conclusions. As pointed by the authors in  \cite{MDSH09}, a lot of published articles in major conferences hide these details, meaning that their experimental results are meaningless.
	\item Usage of deprecated machines, deprecated OS, exotic environment, etc. If we take a research article published five years after the experiences for instance, there is a high chance that the workstations that served the experiences have already died or already changed their behaviour (usury of hardware, software patches, etc.).
\end{itemize}

With the huge amount of published articles in the code optimisation community, with the impressive published speedups, an external reviewer of our community has the right to ask the following naive question: {\it If we combine all the published speedups (accelerations) on the well known public benchmarks since four decades, why don't we observe execution times approaching to zero\,?} This question is justified, and brings a reforming malaise to us. Now, we are asked to be clear about our statistics, some initiatives start to collect published performance data in order to compare them\cite{FuTe:09}. 

The malaise raised by the above question is not a suspicion of a general {\it cheating} in research. We believe that our community is honest in publishing data, but the published observed speedups are sometimes {\it rare} events far from what we could observe if we redo the experiences multiple times. Even if we take an ideal situation where we use exactly the original experimental machines and software, it is too difficult to reproduce exactly the same performance numbers again and again, experience after experience. Usually, published speedups are computed with bias describing pretty {\it rare} events. Frankly, if a computer scientist succeeds in reproducing the performance numbers of his colleagues (with a reasonable error ratio), it would be equivalent to what rigorous probabilists and statisticians call a {\it surprise}.

\subsection{Why Program Execution Times Vary}
What makes a binary program execution time to vary, even if we use the same data input, the same binary, the same execution environment?
\begin{itemize}
	\item Background tasks, concurrent jobs, OS process scheduling;
	\item Interrupts;
	\item Input/output;
	\item Starting loader address;
	\item Branch predictor initial state;
	\item Cache effects;
	\item Non deterministic dynamic instruction scheduler;
	\item Temperature of the room (dynamic voltage/frequency scaling service)
\end{itemize}

One of the reasons of the non-reproducibility of the results is the variation of execution times of the same program given the same input and the same experimental environment. With the massive introduction of multicore architectures, we believe that the variations of executions times will become exacerbated because of the complex dynamic features influencing the execution: threads scheduling policy, synchronisation barriers, resource sharing between threads, hardware mechanisms for speculative execution, etc. Consequently, if you execute a program (with a fixed input and environment) $k$ times, it is possible to obtain $k$ distinct execution times. The mistake here is to assume that these variations are minor, and are stable in general. The variation of execution times is something that we observe everyday, we cannot neglect it. An usual error in the community is to replace all the $k$ execution times by one value, such that the minimum, the mean or the maximum. Doing that would produce {\it sexier} speedups to publish, but does not reflect the reality with fair numbers.

\subsection{Why Don't we Consider the Minimum Execution Time?}
Considering the minimum value of the $k$ observed execution times is unfair because:
\begin{itemize}
	\item nothing guarantees that this minimum execution time is an ideal execution of the program. 
	\item nothing guarantees that this minimum execution time is a consequence of the optimisation technique under study. Maybe this minimum execution time is an accident, or a consequence of dynamic voltage scaling, or anything else.
	\item if this minimal execution time is a rare event, all your statistics describe rare speedups. So, they become non-reproducible easily. 
\end{itemize}
\subsection{What is Inside this Article, What are its Limitations}
We base our reasoning here on common well known results in statistics, especially on some results explained in the book of Raj Jain \cite{Jain:1991:ACS}. We propose a first step towards a rigorous statistical methodology to evaluate program optimisation techniques. This article recalls some common mistakes in performance evaluation, explains which statistics should be used in a particular situation, and provide practical examples. Furthermore, we show how to use the free software called R to compute these statistics \cite{R:2008,R:base}.

Our article is organised to help computer scientists (and of course PhD students) willing to make correct and rigorous statistical study of their code optimisation method. The question is how to convince real experts by statistics, provided a confidence level $\alpha \in ] 0\%, 100\% [$, that your code optimisation technique is really efficient in practice. Section~\ref{sec:speedup} explains when we can decide about a speedup of a program and how we can measure it using $k$ observations of execution times. Having a set of $n$ distinct independent programs (considered as a set of benchmarks), Section~\ref{sec:overal_speedup} explains how to compute an average speedup (while it is a bad idea to synthesise a set of speedups in by a unique average). Getting a speedup (acceleration) inside a sample of $n$ benchmarks does not guarantee you that you can get a speedup on another program. Consequently, Section~\ref{sec:prop_conf_interval} shows how we can estimate the chance that the code optimisation would provide a speedup on a program non belonging to the initial sample of benchmarks used for experiences. 

The limitations of this article are: we do not study the variation of execution times due to changing the program input. We consider real executions, not emulation/simulation nor executions on virtual machines. We also consider a fixed (universal ?) experimental environment. 

\section{Computing a Speedup Factor for a Single Program with a Single Data Input} \label{sec:speedup}
Let $\mathcal{P}$ be an initial program, let $\mathcal{P'}$ be a transformed version after applying the code optimisation technique under study. If you execute the program $\mathcal{P}$ $k$ times, it is possible to obtain $k$ distinct execution times (especially if the program is short): $t_1,\cdots,t_k$.  The transformed program $\mathcal{P'}$ can be executed $m$ times producing $m$ execution times too $t'_1,\cdots,t'_m$. The unit of measure here is the milisecond in general, so we can consider a timing precision in seconds with three digits after the coma. Below is a list of elementary recommendations before starting statistics:
\begin{enumerate}
	\item $\mathcal{P}$ and $\mathcal{P'}$ must be executed with the same data input in {\it similar} experimental environment. The community of code optimisation has not decided yet on the exact semantics of {\it similar}, since many unknown/hidden factors may influence the experiences.
	\item Statistically, it is not necessary that $k=m$. However, it is {\it strongly} recommended that $k \ge 30$ and $m \ge 30$. 30 runs may seem quite prohibitive, but this is the practical limits of the number of observations used in statistics if you want to have a precise Student test that we will explain later. If the number of observations is below 30, computing the confidence intervals of the mean time becomes more complex: we should first check the normality of the distribution (using the normality test of Shapiro-Wilk for instance). If the normality check succeeds, then the test of Student can be applied. Otherwise, the confidence intervals of the mean execution times must be computed using complex bootstrap methods \cite{bootstrap} instead of the test of Student. We highly recommend 30 runs per program to ensure the validity of the Student test. If the program execution time is too large to consider 30 executions, you can do less executions but you should follow the method we just described (either a normality check followed by a Student test, or by using bootstrap methods).
	\item It is important that the repetitive executions of the same program should be independent. For instance, it is not fair to use a single loop around a code kernel that repeat the execution $k$ times. This is because repeating a program $\mathcal{P}$  inside a loop makes them to execute inside the same application. Consequently, the operating system does not behave as if you execute the program $k$ times from the shell. Furthermore, the caches are warmed by the repetitive executions of the code kernels if they belong to the same application. 
	\item Anyway, even if we execute a program $k$ times from the shell, the executions are not necessarily independent, especially if they are executed back-to-back: the seek time of the disk is altered by repetitive executions, some data are cached on the disk by applications and benefit from repetitive executions. Recently, we have been told that branch predictors are also influenced by separate applications: this seems strange, but we should stay careful with hardware mechanisms. As you can see, it is not easy to guarantee $k$ independent executions! 
\end{enumerate} 

\begin{figure}
\begin{center}
\input{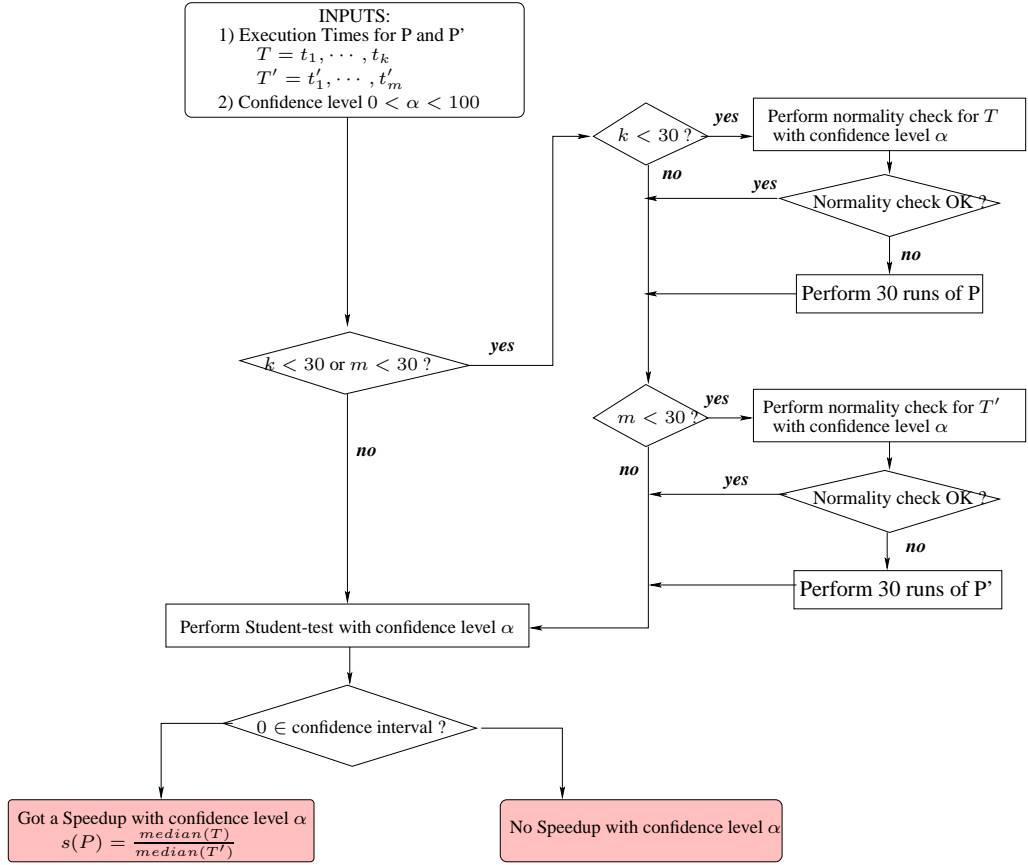}
\end{center}
\caption{Statistical Protocol for Asserting and Computing a Speedup with Confidence Level $\alpha$}
\label{fig:stat_protocole}
\end{figure}

We have remarked a common mistake in computing speedups in presence of program execution time variance: assuming that the variations in execution times are not really a problem, because caused by external factors, these variations may be neglected and smoothed. Consequently, we may be asked to compute the speedup resulted from transforming $\mathcal{P}$ into $\mathcal{P'}$ by using one of the following fractions: $\frac{\min_{i=1,k}t_i}{\min_{j=1,m}t'_j}$, $\frac{\max_{i=1,k}t_i}{\max_{j=1,m}t'_j}$, or  $\frac{\overline{\mu}(\mathcal{P})}{\overline{\mu}(\mathcal{P'})}$. Here, $\overline{\mu}$ is the usual notation of the sample arithmetic mean: $\overline{\mu}(\mathcal{P})=\frac{ \sum_{i=1,k}t_i} {k}$, $\overline{\mu}(\mathcal{P'})=\frac{ \sum_{i=1,m}t'_i}{m}$. If one of the previous speedups is higher than 1, than people conclude victory. The mistake here is to assume that $k$ observed execution times represent any future execution time of the program, even with the same data input. Statistically, we are wrong if we do not consider confidence intervals. To be rigorous, we can follow the four major steps described below to assert a high confidence in the computed speedup. The whole detailed protocol is illustrated in Figure~\ref{fig:stat_protocole}.

\subsection{Step 1: If the Number of Runs is Below 30, Check the Normality}
As said before, if the number of runs is at least 30, you can skip this step. If the number of runs of a program is below 30, we should check if the values $(t_1,\cdots,t_k)$ and $(t'_1,\cdots,t'_m)$ follow a normal distribution. In practice, we can use the Shapiro-Wilk normality test providing a confidence level. The user should fix a confidence level (say $\alpha=95\%$), and the Shapiro-Wilk test can determine (with $1-\alpha=5\%$ chance of error) that the values follow a normal distribution. A later example will show how to practice this using the R software. If the normality check fails, you can either run more executions till 30, or use complex bootstrap method (that we will not explain here).

\subsection{Step 2: Perform a Student Test to Decide if a Speedup Occurs}
The Student test allows to statistically check if all the future executions of the program $\mathcal{P'}$ are faster than the executions of $\mathcal{P}$ with a fixed confidence level $\alpha$ ($0<\alpha<100$). The Student test allows to say that we have $\alpha\%$ of chance that the mean execution time of $\mathcal{P'}$ is faster than the mean execution time of $\mathcal{P}$ by just analysing the $n+m$ observations. This test estimates the confidence interval of the difference between the mean execution times of $\mathcal{P}$ and $\mathcal{P'}$. If the value zero is inside the confidence interval, then the Student test does not guarantee with a confidence level $\alpha$ that the program $\mathcal{P'}$ is faster in average than the program $\mathcal{P}$. That is, if 0 belongs to confidence interval of the Student test, no speedup can be concluded for the program $\mathcal{P}$.

Let $[a,b]$ be the confidence interval computed by the Student test. If $0<a$ then we can say that $\mathcal{P'}$ would be faster in average than $\mathcal{P}$ in $\alpha \%$ of the future executions (considering the same data input and experimental environment). An example is illustrated later.

\subsection{Step 3: If the Student Test Concedes a Speedup, We then Can Measure it}
The speedup factor for the program  $\mathcal{P'}$ can be defined as the fraction between the sample mean times, as follows:
$\frac{\overline{\mu}(\mathcal{P})}{\overline{\mu}(\mathcal{P'})}$. Here, $\overline{\mu}$ is the usual notation of the sample arithmetic mean: $\overline{\mu}(\mathcal{P})=\frac{ \sum_{i=1,k}t_i} {k}$, $\overline{\mu}(\mathcal{P'})=\frac{ \sum_{i=1,m}t'_i}{m}$.
The problem with this definition of speedup is that it is sensitive to {\it outliers}. Indeed, the distributions of the values of $(t_1,\cdots,t_k)$ and $(t'_1,\cdots,t'_m)$ may be biased. 

For the above reasons, we prefer using the median as suggested by \cite{Jain:1991:ACS} instead of the sample mean of the execution times\footnote{Keeping the median execution time is currently used by the SPEC benchmarks for instance}. Consequently, the speedup becomes

$$s(P')=  \frac{\textrm{median}_{i=1,k} t_i} {\textrm{median}_{j=1,m} t'_j}$$

Remember that this speedup has no sense if the Student test fails to determine if 0 is outside the confidence interval. For the remaining part of the article, we note by $m(\mathcal{P})$ and $m(\mathcal{P'})$ the observed median of the execution times of the program $\mathcal{P}$ and $\mathcal{P'}$ resp.

\begin{example}
Let $\mathcal{P}$ be a initial program with its "representative" data input. We are willing to statistically demonstrate with a confidence level $\alpha=95\%$ that an optimisation technique transforms it into $\mathcal{P'}$ and produces benefit in terms of execution speed.  For doing this, I should execute $\mathcal{P}$ and  $\mathcal{P'}$ at least 30 times. For the sake of the example, I consider here only 5 executions for $\mathcal{P}$ and $\mathcal{P'}$. Using the software R, I introduce the values of execution times (in seconds) of $\mathcal{P}$ and $\mathcal{P'}$ as two vectors $T_1$ and $T_2$ resp.

\begin{verbatim}
> library(stats)
> T1<- c(2.799, 2.046, 1.259, 1.877, 2.244)
> T2 <- c(1.046, 0.259, 0.877, 1.244, 1.799)
\end{verbatim}
We must not hurry to conclude and to publish the following result: the resulted speedup for this program is equal to $\min(T1)/\min(T2)=4.86$. Publishing such performance gain (acceleration of factor equal to 4.86) is a statistical mistake. Since we have only 5 observations instead of 30, we should check the normality of the values of $T_1$ and $T_2$ resp. using the test of Shapiro-Wilk.
\begin{verbatim}
> shapiro.test(T1)
	Shapiro-Wilk normality test
data:  T1 
W = 0.9862, p-value = 0.9647
\end{verbatim}
The test of Shapiro-Wilk on the data $T1$ computes here a value $W=0.9862$. In order to say that the test succeeds with confidence level $\alpha$, the value $W$ must be greater  or equal to the $W$ value of the Shapiro-Wilk table (this table can be found on Internet for instance). I use here a confidence level $\alpha=95\%$. The Shapiro-Wilk table for $n=5$ (number of values) and $\alpha=0.95$  indicates the value of 0.986. Consequently, the normality test succeeds for $T1$. Idem for $T2$. 
\begin{verbatim}
> shapiro.test(T2)
	Shapiro-Wilk normality test
data:  T2 
W = 0.9862, p-value = 0.9647
\end{verbatim} 
Since  $W = 0.9862\ge 0.986$, the values of $T_2$ follows a normal distribution with a confidence level of $\alpha=0.95$. It is important to notice here that if the normality test fails for a program ($T_1$ or $T_2$), we must run it at least 30 times.
I can now continue with the Student test to check if $\mathcal{P}'$ is faster than $\mathcal{P}$ with a very high confidence level $\alpha=99\%$.
\begin{verbatim}
> t.test(T1,T2, alternative="greater", conf.level=0.99)
Welch Two Sample t-test
...
99 percent confidence interval:
-0.02574667         Inf 
...
\end{verbatim}
The obtained confidence interval for the difference between the mean execution times is $[-0.02 , +\infty]$. This interval includes 0. Consequently, we cannot assert with $99\%$ confidence level that $\mathcal{P}'$ is faster in average than $\mathcal{P}$. I have the choice by either rejecting the obtained speedup (too hard), or reduce my confidence level. I check with $\alpha=95\%$ instead of $95\%$
\begin{verbatim}
> t.test(T1,T2, alternative="greater", conf.level=0.95)
...
95 percent confidence interval:
 0.3414632       Inf 
...
\end{verbatim}
The confidence interval is $[ 0.34, +\infty]$, it does not include 0. Consequently, we can assert with 95\% confidence level that we obtained a speedup.  In other words, the risk (of error) of not obtaining an acceleration for the future executions is equal to 5\%. The obtained speedup is $s(\mathcal{P})=\frac{m(\mathcal{P})}{m(\mathcal{P'})}=\frac{2.046}{1.046}=1.95$.
\end{example}

\subsection*{Remark: Speedup with Low Confidence Level}
If the confidence level used for the Student test is too low, it is not impossible that we reach a situation where the Student test detects a speedup while the computed speedup is $< 1$. The following example shows that low confidence levels may bring incoherent speedup measure.
\begin{example}
Let take the same previous example with $T_1$ and $T_2$. We apply a Student Test with a confidence level equal to 1\% to ensure that  $\mathcal{P'}$ is slower than  $\mathcal{P}$. In the previous example, we showed the contrary with a confidence level equal to 95\%.
\begin{verbatim}
> t.test(T2, T1, alternative="greater", conf.level=0.01)
...
1 percent confidence interval:
 0.02574667        Inf 
...
\end{verbatim}
As you can see, the test of Student succeeds, so we have 1\% of chance that  $\mathcal{P'}$ is slower than  $\mathcal{P}$. The computed speedup (either by considering the sample mean of the median) is as follows:
\begin{verbatim}
 > mean(T2)/mean(T1)
[1] 0.5110024
> median(T2)/median(T1)
[1] 0.5112414
\end{verbatim}
As you can see, the speedup here is below 1. Is this a contradiction ? No of course, remember that the confidence level of this speedup is only 1\%. 
\end{example}

This section explained how to check with a confidence level $\alpha$ that a code optimisation technique produces a faster transformed program (for a fixed data input and experimental environment).  We also provided a formula for quantifying the speedup. The following section explains how to compute an overall average of speedups of a set of benchmarks. 
\section{Computing the Overall Speedup of a Set of Benchmarks} \label{sec:overal_speedup}
When we implement a code optimisation technique, we are generally asked to test it on a set of benchmarks, not on a unique one. Let $n$ be the number of considered benchmarks. Ideally, the code optimisation technique should produce speedups on the $n$ programs (at least no slowdown) with the same confidence level $\alpha$. Unfortunately, this situation is rare nowadays. Usually, only a fraction of $p$ programs among $n$ would benefit from an acceleration. Let $s( \mathcal{P}_j)$ be the obtained speedup for the program $\mathcal{P}_j$. While this is not correct in statistics, some reviewers ask an average speedup of all the benchmarks. In statistics, we cannot provide a fair average because the programs are different, and their weights are different too. So, asking for an overall speedup for a set of benchmarks will highly bring unfair value. Neither an arithmetic mean, nor a geometric or harmonic mean can be used to synthesise in a unique speedup of the whole set of benchmarks.
 
The arithmetic mean does not distinguish between short and long programs: for instance, having a speedup of  105\% on a program which lasts 3 days must not have the same impact as a speedup of 300\% obtained on a program which lasts 3 seconds. In the former, we save 5\% of 3 days (=216 minutes), while in the latter we save 200\% of 3 seconds (=2 seconds). If we use the arithmetic mean, we would obtain an overall speedup equal to (105+300)/2=202\%, this does not reflect the reality with a fair number.

The geometric mean cannot be applied here because we are not faced to a succession of accelerations on the same program, but to accelerations to distinct programs. The harmonic mean in our case is not meaningful too because the quantity $\frac{1}{s}$ represents also a sort of speedup, so we can provide the same criticism as the arithmetic mean . 

In order to compute $G$ an overall {\it performance gain factor} (not an overall speedup) that represents the weights of the different programs, we can use the following method. The confidence level of this performance gain factor is equal to the minimal value of confidence levels used in the Student tests to validate individual speedups.

First, an interesting question is to decide if we should neglect the $n-p$ programs where no speedup has been validated by the Student test. That is, the performance gain factor is computed for a subset $p$ of programs, not on all the $n$ benchmarks. We believe we neglect the $n-p$ programs that fail in the Student test if we study afterwards (in the next section) the confidence interval of the proportion $\frac{p}{n}$: studying this proportion helps us to decide if the reported overall gain is meaningful. If we decide to include all the $n$ programs for computing the overall  performance gain factor, this is also fair, but the reported gain may be negative since it includes the slowdowns.

Second, we associate a weight $W(\mathcal{P}_j)$ to each program $\mathcal{P}_j$. The general characteristics of a weight function is $\sum_j W(\mathcal{P}_j)=1$. If not, we should normalise the weights so that they sum to 1. The weight of each benchmark can be chosen by the community, by the benchmark organisation, by the user, or we can simply decide to associate the same weight to all benchmarks.  Also, it is legitimate to choose the weight as the fraction between the observed execution time and the sum of all observed  execution times:  $W(\mathcal{P}_j)=\frac{\textrm{ExecutionTime}(\mathcal{P}_j)}{\sum _{i=1,p} \textrm{ExecutionTime}(\mathcal{P}_i)}$. Here we  choose to put $\textrm{ExecutionTime}(\mathcal{P}_j)=m(\mathcal{P}_j)$, ie, the median of all the observed execution times of the program $\mathcal{P}_j$. Someone would argue that this would give more weight on long running time programs: the answer is yes, because what we want to optimise at the end is the absolute execution time, not the relative one.

Third, transforming a program $\mathcal{P}_j$ into $\mathcal{P'}_j$ allows to reduce the execution time by $\textrm{ExecutionTime}(\mathcal{P}_j)-\textrm{ExecutionTime}(\mathcal{P'}_j)$. This absolute gain should not be considered as it is, but should be multiplied by the weight of the program as follows: $g(\mathcal{P}_j)= W({\mathcal{P}_j}) \times ( \textrm{ExecutionTime}(\mathcal{P}_j)-\textrm{ExecutionTime}(\mathcal{P'}_j))$.

Fourth and last, the overall performance gain factor is defined as the fraction between weighted gains and the sum of  weighted initial execution times: $G=\frac{\sum_{j=1,p} g(\mathcal{P}_j)}{ \sum_{j=1,p} W({\mathcal{P}_j}) \times \textrm{ExecutionTime}(\mathcal{P}_j)}$. By simplification, we obtain:
$$G= 1- \frac{ \sum_{j=1,p}  W({\mathcal{P}_j}) \textrm{ExecutionTime}(\mathcal{P}'_j)}{\sum_{j=1,p}  W({\mathcal{P}_j}) \textrm{ExecutionTime}(\mathcal{P}_j)} $$
By definition, the overall gain $G<1$, since the execution times of the optimised programs are hopefully non zero values ($\textrm{ExecutionTime}(\mathcal{P}'_j)\neq 0$).

\begin{example}
Let a program P1 that initially lasts 3 seconds. Assume we succeed to accelerate it with a factor of 300\% with a confidence level $\alpha_1=95\%$. Thus, its new median execution time becomes 1 second. Let P2 be a program that initially lasts 1 hour and has been accelerated with a factor of 105\% with a confidence level $\alpha_2=80\%$. Thus, its new median execution time becomes 3428 seconds. The arithmetic mean of these two speedups is 202.5\%, the geometric mean is 177.48\% and the harmonic mean is 155.56\%. None of these means is suggested for publications as explained before. The weights of the programs P1 and P2 are resp. $W(P1)=3/(3600+3)=0.0008$ and $W(P2)=3600/(3600+3)=0.9991$. The obtained weighted gain for each program is: $g(P1)= 0.001$ and $g(P2)= 171.85$. The overall performance gain factor is then $G=1- \frac{0.0008\times 1 + 0.9991 \times 3428}{0.0008\times 3 + 0.9991 \times 3600} =4.77\%$ and the confidence level is equal to $\alpha=\min(\alpha_1,\alpha_2)=80\%$. 
If we consider that the weights are unit,  $W(P1)=W(P2)=1$, then the overall performance gain factor is then $G=1- \frac{1 + 3428}{3 + 3600} = 4.82\%$ and the confidence level is still equal to $\alpha=\min(\alpha_1,\alpha_2)=80\%$.  As can be remarked, there is not a direct comparison between the overall gain and the individual speedups.
\end{example}

 The following section gives a method to evaluate the quality of a code optimisation method. Precisely, we want to evaluate the chance that a code optimisation technique produces a speedup on a program that does not belong to the initial set of experimented benchmarks.
\section{A Qualitative Evaluation of a Code Optimisation Method}  \label{sec:prop_conf_interval}
Computing the overall performance gain for a sample of $n$ programs does not allow to estimate the quality nor the efficiency of the code optimisation technique. In fact, within the $n$ programs, only a fraction of $p$ benchmarks have got a speedup, and $n-p$ programs got a slowdown.  If we take this sample of $n$ program as a basis, we can measure the chance of getting the fraction of accelerated programs as $\frac{p}{n}$. The higher is this proportion, better would be the quality of the code optimisation. In fact, we want to estimate if the code optimisation technique is beneficial for a large fraction of programs.  The proportion $C=\frac{p}{n}$ has been observed on a sample of $n$ programs. The confidence interval for this proportion (with a confidence level $\alpha$) is given by the equation $ C \mp  r$, where $r=z_{(1+\alpha)/2} \times \sqrt{\frac{C (1-C)}{n}} $. In other words, the confidence interval of the proportion is equal to $I=\left [ C -  r, C +  r \right ]$. Here, $z_{(1+\alpha)/2}$ represents the value of the $(1+\alpha)/2$ quartile of the unit normal form. This value is available in a known table (table A.2 in \cite{Jain:1991:ACS}). The confidence level $\alpha$ is equal to the minimal value of confidence levels used in the Student tests to validate individual speedups. We should notice that the previous formula of the confidence interval of the proportion $C$ is valid only if $n.C\ge 10$. If $n.C < 10$, computing the confidence interval becomes too  complex according to  \cite{Jain:1991:ACS}. 
\begin{example}
Having $n=30$ benchmarks, we obtained a speedup on only $p=17$ cases. We want to compute the confidence interval for the proportion C=17/30=0.5666 with a confidence level $\alpha=0.9=90\%$.  The quantity $n.C=17 \ge 10$, I can then easily estimate the confidence interval of $C$ using the R software as follows. 
\begin{verbatim}
> prop.test(17, 30, conf.level=0.90)
...
90 percent confidence interval:
 0.4027157 0.7184049 
...
\end{verbatim} 
 \end{example}
The above test allows us to say that we have 90\%  of chance that the proportion of accelerated programs is between $40.27\%$ and $71.87 \%$. If this interval is too wide for the purpose of the study, we can reduce the confidence level as a first straightforward solution. For instance, if I consider $\alpha=50\%$, the confidence interval of the proportion becomes $[49.84\%, 64.23\%]$. Or, if we do not want to reduce the confidence level, we need to do more experiences on more benchmarks. 

The next formula gives the minimal number $n$ of benchmarks requested if we want to estimate the confidence interval with a precision equal to $r\%$ with a confidence level $\alpha$: $$n\ge (z_{(1+\alpha)/2})^2 \times  \frac{C(1-C)}{r^2}$$
\begin{example}
 In the previous example, we have got an initial proportion equal to $C=17/30=0.5666$. If I want to estimate the confidence interval with a precision equal to 5\% with a confidence level of 95\%, I put $ r= 0.05$ and I read in the quartiles tables $z_{(1+0.95)/2}=z_{0.975}=1.960$. The minimal number of benchmarks to observe is then equal to: $n \ge 1.960^2 \times \frac{0.566\times (1-0.566)}{0.05^2}=377.46$. We need to experiment 378 benchmarks in order to assert that we have 95\% of chances that the proportions of accelerated programs are in the interval $0.566 \mp 5 \%$.

The discussion that we can have here is on the quality or on the representativeness of the sample of $n$ benchmarks. This is outside the scope of the paper! Until now, we do not know what does a set of representative programs means.

\end{example}

\section{Conclusion}
Program performance evaluation and their optimisation techniques suffer from the disparity of the published results. It is of course very difficult to reproduce exactly the experimental environment since we do not always know all the details or factors influencing it. This article treats a part of the problem by recalling some principles in statistics allowing to consider the variance of program execution times. The variance of program execution times is not a chaotic phenomena to neglect or to smooth; we should keep it under control and incorporate it inside the statistics we publish. This would allows us to assert with a certain confidence level that the results we publish are reproducible under similar experimental environment. 

Using simulators instead of real executions provide reproducible results, since simulators are deterministic: usually, simulating a program multiple times should always produce the same performance numbers. This article assumes that the observations have been done on the physical machine  not by simulation. If the physical machine does not exist, the observations based on simulation cannot be studied exactly with the methods described in this article. The study should more be concentrated on the statistical quality of the simulator. As far as we know, it does not exist yet a simulator that has been rigorously validated by statistics as described in \cite{Jain:1991:ACS}. Usual error ratios reported by simulators are not sufficient alone to judge about their quality.

This article does not treat performance evaluation with multiple data inputs of a program. In fact, the speedups defined in this article are computed for a unique set of data input. Experimenting multiple sets of data input to measure a speedup is let for a future work.

We conclude with a short discussion about the confidence level we should use in this sort of statistical study. Indeed, there is not a unique answer to this crucial question. In each context of code optimisation we may be asked to be more or less confident in our statistics. In the case of hard real time applications, the confidence level should be high enough (more than 95\% for instance), requiring more experiments and benchmarks. In the case of soft real time applications (multimedia, mobile phone, GPS,  etc.), the confidence level can be more than 80\%. In the case of desktop applications, the confidence level should not be necessarily high. In any case, the used confidence level for statistics must be declared for publication.

\section*{Acknowledgement}
We would like to thank Sebastien \textsc{Briais} from the University of Versailles Saint-Quentin en Yvelines for his helpful remarks to improve this document.
\end{document}